\documentclass[aps,twocolumn,floatfix]{revtex4}
\usepackage{epsfig}
\newcommand{\be}{\begin{equation}}
\newcommand{\ee}{\end{equation}}

\begin{document}

\title{Heat Conduction and Entropy Production in a One-Dimensional Hard-Particle Gas}

\author{Peter Grassberger, Walter Nadler, and Lei Yang}

\affiliation{John-von-Neumann Institute for Computing, Forschungszentrum J\"ulich,
D-52425 J\"ulich, Germany}

\date{\today}

\begin{abstract}
We present large scale simulations for a one-dimensional chain of hard-point particles 
with alternating masses. We correct several claims in the recent literature based on 
much smaller simulations. Both for boundary conditions with two heat baths at different 
temperatures at both ends and from heat current autocorrelations in equilibrium we 
find heat conductivities $\kappa$ to diverge with the number $N$ of particles. 
These depended very strongly on the mass ratios, and extrapolation to 
$N\to\infty$ resp. $t\to\infty$ is difficult due to very large finite-size and
finite-time corrections. Nevertheless, our 
data seem compatible with a universal power law $\kappa \sim N^\alpha$ with $\alpha
\approx 0.33$. This suggests a relation to the 
Kardar-Parisi-Zhang model. We finally show that the hard-point gas with periodic
boundary conditions is not chaotic in the usual sense and discuss why the system,
when kept out of equilibrium, leads nevertheless to energy dissipation and entropy
production.
\end{abstract}

\maketitle

Low-dimensional systems are special in many ways. Second order phase transitions 
have anomalous exponents, chemical reactions do not follow the mass action
law \cite{oz78}, hydrodynamics breaks down due to divergent transport coefficients 
caused by long time tails \cite{aw67}, and electrons 
in disordered systems are localized \cite{a58}. A last item in this list  
is the divergence of heat conductivity \cite{llp01} in $\le 2$ spatial dimensions.

For ordered (periodic) harmonic systems it is well known that all transport 
coefficients are infinite due to the ballistic propagation of modes. Thus one 
needs either disorder or nonlinear effects in order to have finite conductivity 
$\kappa$. For electric conductivity, disorder in 1-d leads to zero conduction. 
The main difference between heat and (electronic) charge conduction is that there 
is no background lattice in the former, i.e. translation invariance is not broken 
even if the system is disordered.
Of course one can study the electronic contribution to heat conduction, but 
experimentally one never can neglect the ionic contribution \cite{foot0}.
Thus one has always soft (Goldstone) modes in heat conduction.
These soft modes are not localized by disorder \cite{mi70}, and they are not 
affected by nonlinearities. Thus they propagate essentially freely. In high 
dimensions this has no dramatic consequences. But in low dimensions they become 
important and lead to the above mentioned divergence. More precisely, one expects 
a power divergence $\kappa \sim N^\alpha$ in $d=1$ and a logarithmic divergence 
in $d=2$. Simulations and calculations using the Green-Kubo formula give 
$\alpha \approx 0.35$ to 0.45 \cite{llp01}. It is not clear whether the slight
discrepancies found between different models are true or artifacts. Theoretically,
one would of course prefer a universal value.

There are some exceptions to this general scenario. Apart from models 
with external potentials and broken translation invariance, the best known 1-d
model with finite $\kappa$ is the rotor model of \cite{glpv00,gs00}. Here 
very fast rotors effectively decouple from their less fast neighbours. Thus 
very steep jumps of temperature are effectively stabilized and act as barriers
to energy transport.

Another model which was recently claimed to have $\kappa$ finite \cite{ghn01}
is the 1-d hard point gas with alternating masses. The same gas with all 
particles having the same mass is trivial (a collision between particles 
is indistinguishable from the particles just passing through each other 
undisturbed), and perturbations just propagate ballistically, leading thus to 
infinite $\kappa$ (i.e., no temperature gradients can build up inside the gas). 
To break this integrability, it is sufficient 
to use alternating masses: every even-numbered particle has mass $m_1$, and 
every odd has $m_2 = r m_1$ with $r>1$ \cite{ghn01,mm83,a93,acg97,h99,a01,a02}. 

The arguments given above for a divergence of $\kappa$ with $N$ hold also
for the hard point gas. It is obviously nonlinear, sound waves dissipate,
it is translation invariant, and there is no obvious special feature as in the 
rotor model. Indeed, prior to \cite{ghn01},
heat conduction has been studied in this system by means of simulations in
\cite{h99,a01}. In these papers it was claimed that $\kappa$ diverges. But 
the simulations of \cite{h99} had presumably low statistics (very few details
were given), while the simulations of \cite{a01} are obviously not yet in the 
scaling regime ($N$ is too small) and are compatible with $\kappa\to const$ 
for $N\to\infty$. In any case, the exponent $\alpha$ suggested by the simulations
of \cite{a01} is $\le 0.22$, much smaller than for all other models.

In view of this confusing situation, and suspecting that the simulations of 
\cite{h99,a01,ghn01} were not done most efficiently, we decided to make 
some longer simulations. 

We followed \cite{a01} in setting $m_1=1$ and using Maxwellian heat baths at 
the ends with $T_1=2, T_2=8$ (i.e., after hitting the end, a particle is reflected
with a random velocity distributed according to $P(v) = \Theta(\pm v) mv/T \exp(
-mv^2/2T_{1,2})$. The heat baths sit at $x=0$ and $x=N$, i.e. the 
density of the gas is 1. When an even particle with velocity $v_1$ collides
with an odd one with velocity $v_2$, their velocities after the collision are
\begin{equation}
   v_1' = {(1-r)v_1 + 2rv_2\over (1+r)}\ ,\quad v_2' = {2v_1+(r-1)v_2\over (1+r)}\;.
                                                          \label{v-update}
\end{equation}
Between two
collisions, the particles propagate freely. Thus a fast simulation algorithm 
is event-driven: For each particle $i$ we remember its velocity, the time $t_i$ 
of its last collision (initially, $t_i$ is put to zero), and the position 
$x_i$ it had at that time. In addition, 
we maintain a list of future collision times $\tau_i$ for each neighboring pair 
$(i,i+1)$ (here the walls are treated formally as particles with 
$v_0=v_{N+1}=0,\; x_0=0,\; x_{N+1}=N$). The system is evolved by searching 
the smallest $\tau_i$, updating the triples $(t_i,x_i,v_i)$ and 
$(t_{i+1},x_{i+1},v_{i+1})$, and calculating the new future collision
times $\tau_{i-1}$ and $\tau_{i+1}$ (the new $\tau_i$ is infinite). Since 
the list $\{\tau_i\}$ is essentially
a priority queue \cite{s90}, we use for it the appropriate data structure of a 
{\it heap} \cite{s90}. Using heaps,
searching for the next collision takes a CPU time $O(\ln N)$. In comparison,
a naive search would take $O(N)$. This allowed us to make much larger simulations
than in previous works. Our largest systems contained 16383 particles and were 
followed for $> 10^{12}$ collisions. In spite of this, we had to 
start with carefully tailored initial configurations to keep transients short. 
When obtaining statistics one should not forget that measurements should {\it not} 
be made after a fixed number of collisions, but at fixed intervals in real time.
The correctness of the results and the absence of transients were checked 
by verifying that the energy density is constant, as proven in \cite{a01}.

Before presenting results, let us discuss the expected dependence on the mass 
ratio $r$. For $r\to 1$, equilibration becomes slow (it takes a long time until 
a fast particle is slowed down to average speed), but perturbations propagate
ballistically. Thus a perturbation will be damped out slowly at first, but it 
will have no long time tails and is damped exponentially. In the other extreme,
for $r\to\infty$ the light particles bounce 
between pairs of heavy ones which are hardly perturbed. Thus, if a heavy 
particle is perturbed, we have a situation very similar to the one for $r\to 1$.
If a light particle is perturbed, its energy is soon given to its two nearest
heavy neighbours, which then behave again as for $r\approx 1$. In contrast, 
in the intermediate region $1\ll r \ll \infty$ we expect the perturbation 
to spread non-ballistically for all times. It is in this regime that we expect the 
fastest convergence to asymptotic behaviour, both with respect to time and to 
$N$.

\begin{figure}
\psfig{file=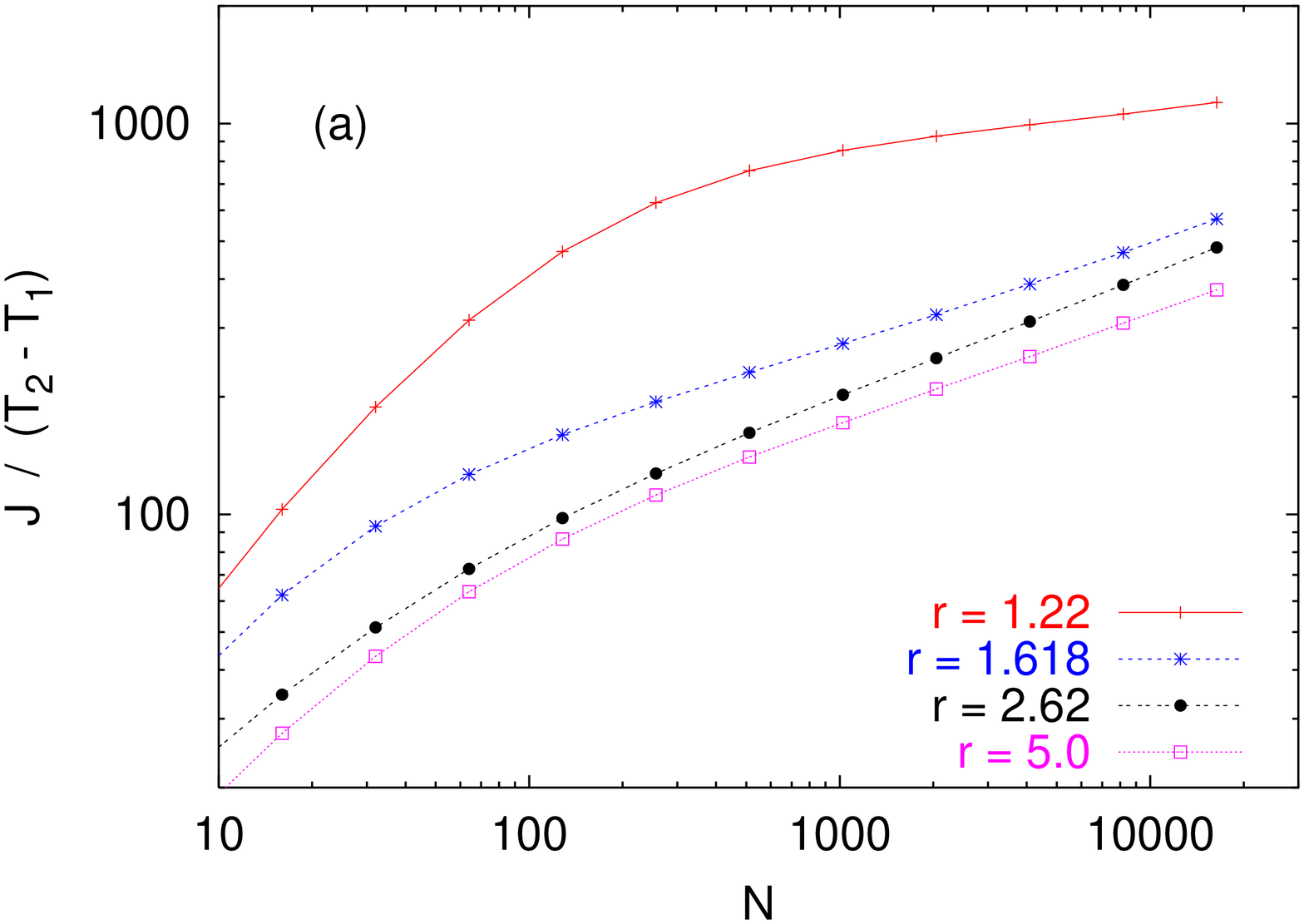,width=5.8cm,angle=270}
\psfig{file=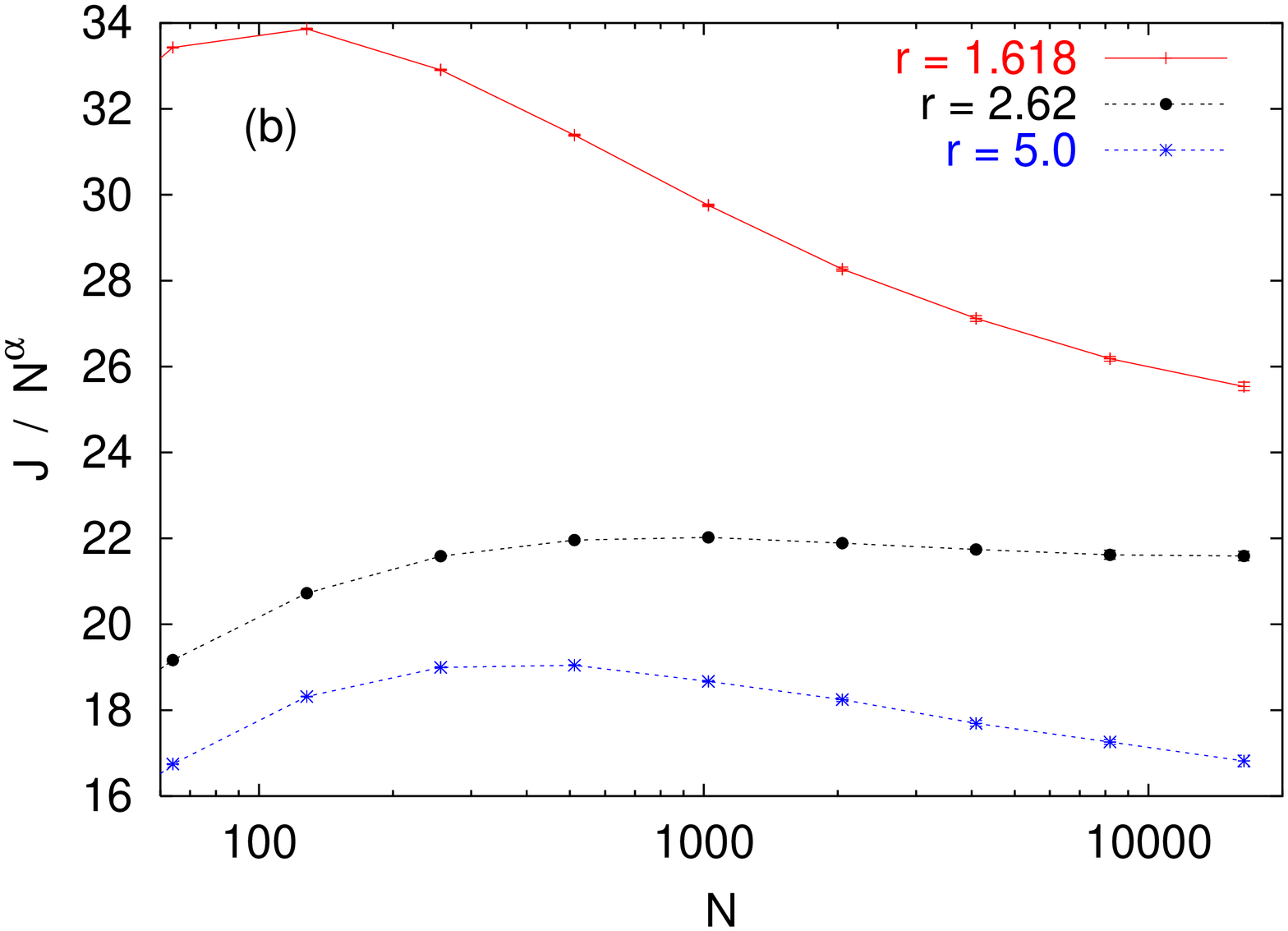,width=5.4cm,angle=270}
\caption{(a): Log-log plot of $J/(T_2-T_1)$ versus $N$ for four values of $r$. 
  Statistical errors are always smaller than the data symbols. (b): Part of the 
  same data divided by $N^\alpha$ with $\alpha=0.32$, so the y-axis is much 
  expanded.}
\label{fig1.ps}
\vglue -4pt
\end{figure}

In Fig.1a we show $\kappa$, defined as total energy flux $J = \sum_i m_i v_i^3/2$
divided by $\Delta T$, versus $N$, for four values of $r$. The value $r=1.22$ is 
in the small-$r$ region and was studied most intensively in \cite{a01}. The value
$r=2.62$ is near the center of the intermediate regime, while $r=5$ is clearly
above it. Finally, $r=1.618 =(1+\sqrt{5})/2$ was chosen because it was used in 
\cite{ghn01}, not because of its irrationality (problems with ergodicity related 
to rational values of $r$ exist only for very small $N$ \cite{a93,acg97}). 

A power law would give a straight line with slope $\alpha$ in Fig.1a. None of the 
four curves is really straight. For small $N$ the curve for $r=1.22$ agrees 
perfectly with Fig.3 of \cite{a01} (which extends only to $N=1281$).
It shows the strongest curvature (in agreement with the above discussion), and 
the small-$N$ data alone would suggest a cross-over to $\kappa = const$. But this 
curvature stops for large $N$ and a closer look shows that the 
slope {\it increases} for $N > 8000$. The same is true also for the other 
curves: They all bend down for small $N$ but veer up for larger systems (Fig.1b). 
This is most clearly seen for $r=1.618$ and 2.62. It is less clear for $r=5$, but 
the most rational expectation is that also this curve will have the same slope for 
$N\to\infty$. Our best estimate $\alpha = 0.32 {+0.03\atop -0.01}$ has 
asymmetric errors because we do not know how much more the curves will bend 
upwards for very large $N$.

\begin{figure}
\psfig{file=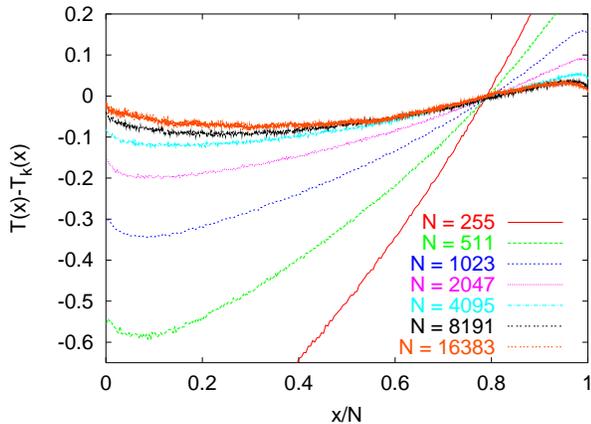,width=5.8cm,angle=270}
\caption{$T(x)-T_k(x)$ against $x/N$ for $r=1.22$, where $T_k(x) = [T_1^{2/3} - 
(T_1^{2/3}-T_2^{2/3})x/N]^{2/3}$ is the temperature profile according to kinetic 
theory. In order to reduce statistical fluctuations, we averaged in the curves 
for $N\ge 1023$ over three successive values of $x$.}
\label{fig2.ps}
\vglue -3pt
\end{figure}

\begin{figure}
\psfig{file=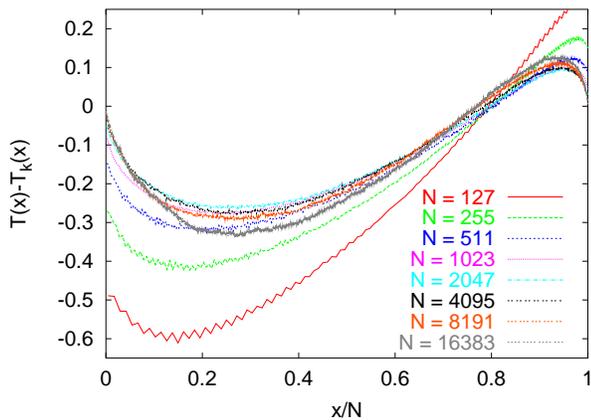,width=5.8cm,angle=270}
\caption{$T(x)-T_k(x)$ against $x/N$ for $r=1.618$.}
\label{fig3.ps}
\vglue -3pt
\end{figure}

The rescaled temperature profiles for $r=1.22$ are shown in Fig.2. To verify
the claim of \cite{a01} that $T(x)$ approaches the profile $T_k(x)$ predicted by 
kinetic theory with an inverse power of $N$, i.e. $T(x)-T_k(x) \sim N^{-0.67}$,
we plot $T(x)-T_k(x)$ against $x/N$. For $N<2000$ we see indeed this 
convergence in perfect agreement with \cite{a01}, but not for $N>2000$. Instead, 
it seems that $T(x)-T_k(x)$ remains different from zero for $N\to\infty$.
The analogous results for $r=1.618$ are shown in Fig.3. In that case, the scaling 
observed in \cite{a01} is confined to very small $N$, not shown in the figure.
The fact that $T(x)-T_k(x)$ remains finite for $N\to\infty$ is now obvious.
In contrast to a conjecture
in \cite{ghn01}, the temperature profile also does not become linear for large $N$.
All results for $r=1.618$ are qualitatively also true for $r=2.62$ (not shown).

These results are easily understood. For $r=1.22$ 
we are in the small-$r$ regime. This explains the slow convergence of $\alpha$
with $N$ and the weakness of long time tails, manifested in the agreement with
kinetic theory. Only at very large $N$ we do see the correct 
asymptotics. For $r=1.618$ and 2.62 we are no longer in this regime, the long time
tails are stronger, and the disagreement with kinetic theory is more obvious.

In addition to systems driven by thermostats at different temperatures, we also 
studied systems in equilibrium with periodic boundary conditions. Here
the Green-Kubo formula allows $\kappa$ to be calculated from an integral 
over the heat current autocorrelation $\langle J(t)J(0) \rangle$ \cite{llp01}. 
In Fig.4 we show this, after suitable normalization and after multiplication by
a power of $t$ which makes it constant for large $N$ and $t$. We see strong 
oscillations with periods $\propto N$ which reflect the dominance of (damped)
sound waves with a fixed velocity of sound (see also \cite{a02}). 
They were mistaken for statistical
fluctuations in \cite{ghn01}, showing clearly that the simulations of \cite{ghn01} 
have not reached the asymptotic regime in contrast to what the authors assumed. 
Our data suggest that $\langle J(t)J(0) \rangle \sim t^{-0.66}$ for large $N$ 
with a cut-off at $t\propto N$,
giving $\alpha = 0.34$ in perfect agreement with our previous estimate.

A 1-d hard particle gas should be described macroscopically by hydrodynamics,
i.e. by the Burgers resp. Kardar-Parisi-Zhang (KPZ) equation \cite{kpz}. If 
we assume that heat diffusion scales like diffusion in KPZ, we might expect 
$\alpha = 1/3$ in agreement with our numerics. But we should warn that 
{\it particle} spreading in the 1-d hard particle gas is {\it not} 
superdiffusive (unpublished data; for $r=1$ see \cite{jepsen}), 
so the relation with KPZ is not trivial.

\begin{figure}
\psfig{file=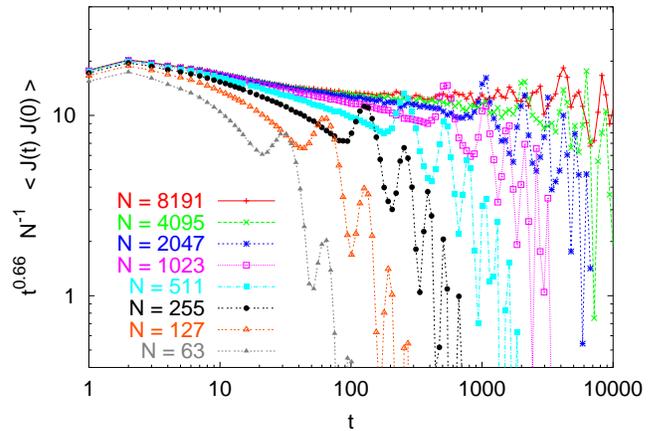,width=6.0cm,angle=270}
\caption{Total heat current autocorrelation, $t^{0.66} N^{-1}$ $\langle J(t)J(0) 
   \rangle$ for $r=2.2$ and $T=2$. Total momentum is $P=0$.}
\label{fig4.ps}
\end{figure}

Finally, we measured also the propagation of infinitesimal perturbations.
Similar simulations were also made in \cite{ghn01},
but there the perturbations were added to the ground state ($E=0$, all particles are 
at rest). In contrast, we perturbed equilibrium states, i.e. we performed a 
standard stability analysis as used e.g. to estimate Lyapunov exponents. Indeed, we
only followed the perturbation in velocity space, not in real space. More precisely, 
after having run the system long enough to have eliminated transients, we chose
a tangent vector
\begin{equation}
   (\delta v_i(0),\delta x_i(0)) = (\delta_{i,0},0)                  \label{tang}
\end{equation}
and iterated, together with the system itself, its linearized variational equations. 
Notice that $\delta v_i(t)$ are independent of the positions perturbations 
$\delta x_i(t)$ for nearly all times (whenever there is no collision), thus it is 
possible and legitimate to solve the variational equations for velocities only.
After the evolution of the $\delta v_i(t)$ has been followed for a given time 
$t_{\rm max}$, $\delta v_i(0)$ is again chosen as 
$\delta_{i,0}$, and the integration is continued.

\begin{figure}
\psfig{file=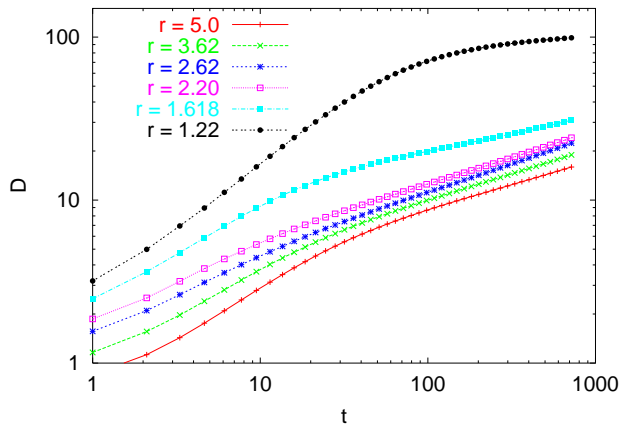,width=5.9cm,angle=270}
\caption{Effective diffusion coefficients for various values of $r$. Statistical
errors are smaller than the sizes of the symbols. Temperature is $T=2$, total 
momentum is $P=0$.}
\label{fig5.ps}
\vglue -3pt
\end{figure}

It is easy to prove that that the 1-d hard point gas is not chaotic in the usual 
sense. Consider the weighted $L_2$ norm of the perturbation,
$||\delta v(t)||_2 = [\sum_{i=1}^N m_i (\delta v_i(t))^2]^{1/2}$.
Since the $\delta v_i(t)$ change during a scattering according to the same 
Eq.(\ref{v-update}) as the velocities $v_i(t)$ themselves, energy conservation leads
to $||\delta v(t)||_2 \equiv 1$. Indeed the absence of chaos is quite obvious since 
there is no local instability. It seems 
to contradict a widespread folklore that dissipation and entropy production are
tightly related to chaos (which sometimes is true; e.g. \cite{gd99}, page 231) -- 
although it is also appreciated that this might not be always the case \cite{dc00}.

One solution to this puzzle is the observation \cite{gs99}, going 
back to work by Wolfram on cellular automata \cite{w84}, that the notion of chaos in 
systems with infinitely many degrees of freedom is ambiguous and is not necessarily 
related to any {\it local} instability. In a spatially extended system it makes 
perfect sense to use a norm which, in contrast to the $L_2$ norm used above, puts 
most weight on near-by regions and exponentially little weight on far-away regions.
With such a definition, the norm of a perturbation moving towards (away from) the 
observer with constant velocity will increase (decrease) exponentially.
More generally, also perturbations spreading 
diffusively will lead to an increase of the uncertainty about the local state for a 
short-sighted observer. For the 1-d hard point gas this means that there is no need 
for any local instability to generate dissipation, local thermal equilibrium, and 
mixing. In a non-equilibrium case entropy {\it flow} is provided by the stochastic 
thermostats at the ends, while (coarse-grained) entropy is {\it produced} by the 
diffusive propagation of perturbations.

Indeed, the situation is not quite as simple due to the divergence of $\kappa$ with
$N$ which suggests that perturbations propagate super-diffusively. For a more 
direct proof, we show in Fig.5 their effective diffusion coefficient, i.e. the 
average squared distances divided by $t$ (notice that $\sum m_i (\delta v_i(t))^2=1$),
\begin{equation}
    D \equiv \langle X^2\rangle / t = t^{-1} \sum_{i=1}^N \;i^2 m_i (\delta v_i(t))^2\;.
\end{equation}
For diffusive spreading, this would be constant. Instead, it increases with $t$ for all
$r$. This increase does not follow a pure power law, but again deviations from a 
power law are strongest for very small and very large $r$. For moderately large $t$ 
they would suggest a crossover to normal diffusion ($D=const$), but for all $r>1.6$ 
this is ruled out by the values at very large $t$. Again it is hard to estimate the
asymptotic behaviour precisely, because the curves bend upward for large $t$. If we 
assume $D\sim t^\beta$, we obtain a precise lower bound $\beta > 0.36\pm 0.01$, but 
only a poor upper bound which just excludes ballistic behaviour.

In summary, we have given compelling evidence that heat conduction in the 1-d hard
point gas shows the anomalous divergence with system size expected for any generic 
1-d system, in spite of strong finite-size and finite-time effects. This ``normal" 
anomalous behaviour holds in spite of the fact that the system is not chaotic in
the usual sense, proving again that chaos in the form of local instabilities is 
not needed for mixing behaviour and dissipation. Finally, we have discussed a possible 
connection to KPZ scaling.

We are indebted to Roberto Livi and Antonio Politi for very helpful correspondence.
W. N. is supported by the DFG, Sonderforschungsbereich 237.

\end{document}